\documentclass[12pt]{article}
\usepackage{authblk}
\usepackage{fullpage}
\usepackage{amssymb,amsmath}
\usepackage[utf8x]{inputenc}
\usepackage[T1]{fontenc}
\usepackage{siunitx}
\usepackage[version=3]{mhchem}
\usepackage{mathtools}
\usepackage{gensymb}
\usepackage[numbers]{natbib}
\usepackage[dvipsnames]{xcolor}
\usepackage{threeparttable, tablefootnote}

\usepackage{pdflscape}
\usepackage{afterpage}

\newcommand{\AB}[1]{\textcolor{black}{#1}}

\usepackage{setspace}
\usepackage[unicode=true]{hyperref}
\hypersetup{breaklinks=true,
            bookmarks=true,
            colorlinks=false,
            pdfborder={0 0 0}}
\usepackage{graphicx}
\graphicspath{{figures/}}
\makeatletter
\def\ScaleWidthIfNeeded{%
 \ifdim\Gin@nat@width>\linewidth
    \linewidth
  \else
    \Gin@nat@width
  \fi
}
\def\ScaleHeightIfNeeded{%
  \ifdim\Gin@nat@height>0.9\textheight
    0.9\textheight
  \else
    \Gin@nat@width
  \fi
}
\makeatother
\setkeys{Gin}{width=\ScaleWidthIfNeeded,height=\ScaleHeightIfNeeded,keepaspectratio}%

\newcommand{\farcs}{\mbox{\ensuremath{.\!\!^{\prime\prime}}}}%

\title{An inherited complex organic molecule reservoir in a warm planet--hosting disk}

\author[1]{Alice S. Booth$^{*}$}
\affil[1]{Leiden Observatory, Leiden University, 2300 RA Leiden, the Netherlands}

\author[2]{Catherine Walsh}
\affil[2]{School of Physics and Astronomy, University of Leeds, Leeds LS2 9JT, UK}

\author[1]{Jeroen Terwisscha van Scheltinga}

\author[1,3]{Ewine F. van Dishoeck}
\affil[3]{Max--Planck--Institut f\"{u}r Extraterrestrishe Physik, Gie{\ss}enbachstrasse 1, 85748 Garching, Germany}

\author[2]{John D. Ilee}

\author[1,4]{Michiel R. Hogerheijde}
\affil[4]{Anton Pannekoek Institute for Astronomy, University of Amsterdam, the Netherlands}

\author[5,6]{Mihkel Kama}
\affil[5]{Department of Physics and Astronomy, University College London, Gower Street, London, WC1E 6BT, UK}
\affil[6]{Tartu Observatory, University of Tartu, Observatooriumi 1, 61602 T\~{o}ravere, Tartumaa, Estonia}

\author[7]{Hideko Nomura}
\affil[7]{National Astronomical Observatory of Japan, 2--21--1 Osawa, Mitaka, Tokyo 181--8588, Japan}

\date{}

\begin{document}

\maketitle

\small $^{*}$Corresponding author email - abooth@strw.leidenuniv.nl

\section*{Abstract}

Quantifying the composition of the material in protoplanetary disks is paramount to determining the potential for exoplanetary systems to produce and support habitable environments. A key complex organic molecule (COM) to detect is methanol (\ce{CH3OH}). \ce{CH3OH} primarily forms at low temperatures via the hydrogenation of CO ice on the surface of icy dust grains and is a necessary basis for the formation of more complex species like amino acids and proteins. We report the detection of \ce{CH_3OH} in a disk around a young, luminous A--type star \AB{HD~100546}. This disk is warm and therefore does not host a significant CO ice reservoir. We argue that the \ce{CH3OH} cannot form in situ, and hence, this disk has likely inherited COMs rich ice from an earlier cold dark cloud phase. This is strong evidence that at least some of the organic material survives the disk formation process and can then be incorporated into forming planets, moons and comets. Therefore, crucial pre--biotic chemical evolution already takes place in dark star--forming clouds.

\section*{Main Text}

\subsection*{Introduction}

Young stars are surrounded by a disk of dust, gas, and ice that will ultimately form an exoplanetary system composed of planets, moons, asteroids, and comets. The potential for these exoplanetary systems to produce and support habitable environments is set by the composition of the material in the parent protoplanetary disk. 
So far, the Earth is the only environment in the Solar System within which life is known to have begun and thrived. 
Comet impacts are thought to have played a key role in this by delivering volatile organic--rich material thus seeding the surface with the basic ingredients for life \citep{doi:10.1146/annurev-astro-091918-104409}.
\AB{Biologically significant molecules for life, e.g., glycine,
and the precursors methylamine and ethylamine, have been detected in comet 67P \citep{Altwegge1600285}.}
The icy moons that orbit Jupiter and Saturn have been earmarked as potential habitats for extraterrestrial life due to the presence of sub--surface liquid oceans \citep{2010SSRv..153..511R}. Also, New Horizons revealed that Kuiper Belt Object (KBO) 486598 (Arrokoth) has a surface rich in organic material including methanol (\ce{CH3OH}) \citep{Grundyeaay3705}. 

Assessing the potential habitability of forming extra--terrestrial planetary systems remains challenging. One way to assess this is by measuring the inventory of organic material present during planet formation. 
Complex organic molecules (COMs) bridge the gap in complexity between the \AB{two or three atom} 
molecules that are typically found in space, with those molecules that are needed for life, such as amino acids and proteins. 
COMs are found to be ubiquitous in the warm surroundings of forming stars where they are seen in the gas phase, after sublimation following formation in CO rich ice in cold dark clouds \citep[][]{doi:10.1146/annurev-astro-032620-021927}. However, it is still uncertain if this organic ice reservoir survives the assembly of the protoplanetary disk \citep[][]{2009A&A...495..881V, 2014MNRAS.445..913D}. 
Recently, observations of cold disks with the Atacama Large Millimeter/submillimeter Array (ALMA) have resulted in the first
detections of simple members of some key functional groups in organic chemistry: acetonitrile \citep[\ce{CH3CN}, a simple nitrile;][]{2015Natur.520..198O,2018ApJ...859..131L}, methanol \citep[\ce{CH3OH}, an alcohol;][]{2016ApJ...823L..10W}, and formic acid \citep[\ce{HCOOH}, a carboxylic acid;][]{2018ApJ...862L...2F}.  
These detections demonstrate that relatively complex molecules are present at the epoch of planet formation, but do not yet provide firm constraints on their origin: in situ formation versus inheritance from the cold cloud phase. 

A COM of particular significance is \ce{CH3OH}. This COM is only formed efficiently on the surfaces of very cold dust grains ($\lesssim 20$~K) 
via the hydrogenation of CO ice \citep[][]{2009A&A...505..629F}.  
Laboratory experiments have revealed \ce{CH3OH} as a feedstock for building molecules of higher complexity, including simple esters, ketones, and aldehydes  \citep[][]{2009A&A...504..891O}. 
Hence, the presence of gas--phase methanol is a key indicator that larger, more complex, molecules could also be present.
The temperature structures of protoplanetary disks strongly depend on the luminosity of the host star.  
\AB{The typical luminosities of Herbig~Ae stars are of order 10-100 solar luminosities compared to 0.01-10.0 solar luminosities for T-Tauri stars \citep{Dunham_2015, 2020MNRAS.493..234W}.} Thus, disks around intermediate--mass Herbig~Ae/Be stars will have a substantially lower fraction of disk mass at the coldest temperatures ($\lesssim 20$~K) than their cooler Sun--like (T--Tauri) counterparts. Oxygen-bearing COMs have so far only been detected in disks around T~Tauri stars.
This means that disks around Herbig Ae/Be stars are expected to be poor in \ce{CH3OH}. 
This is supported by the non--detection of \ce{CH3OH} with deep ALMA observations towards the otherwise molecule--rich Herbig Ae/Be disks around HD~163296 and MWC~480 \citep[][]{2019A&A...623A.124C,2020ApJ...893..101L} as well as detailed astrochemical models that predict a negligible reservoir of CO ice in these systems \citep{2018A&A...616A..19A, 2018A&A...618A.182B}.

We report the detection of this key COM, \ce{CH3OH}, in a disk around the Herbig Ae/Be star, HD~100546. 
HD~100546 is a intermediate--aged ($4.79^{+1.82}_{-0.22}$~Myr) Herbig Be star ($2.18^{+0.02}_{-0.17}$~$\mathrm{M_{\odot}}$; $T_\mathrm{eff} = 10,000$~K) at a distance of 320 light years ($110^{+1}_{-1}$~pc) \citep{2020MNRAS.493..234W}. 
The disk around HD~100546 is warm and gas rich and has been well observed at multiple wavelengths with compelling evidence for two giant planets embedded in the disk at $\approx$10 and $\approx$60~au \citep{2014ApJ...791L...6W, 2013ApJ...766L...1Q, 2014ApJ...791..136B}. 
Our new data uncover a previously unknown reservoir of circumstellar complex organic molecules and show that disks around Herbig Ae/Be stars are \emph{at least} as chemically complex as their Sun--like counterparts. 
Since Herbig Ae/Be disks are warm and cannot form \ce{CH_3OH} in situ, the detection of this COM provides strong evidence for inheritance. 


\subsection*{Results}

\begin{figure*}
\centering
\includegraphics[trim={0cm 0cm 0cm 0cm},clip,width=\hsize]{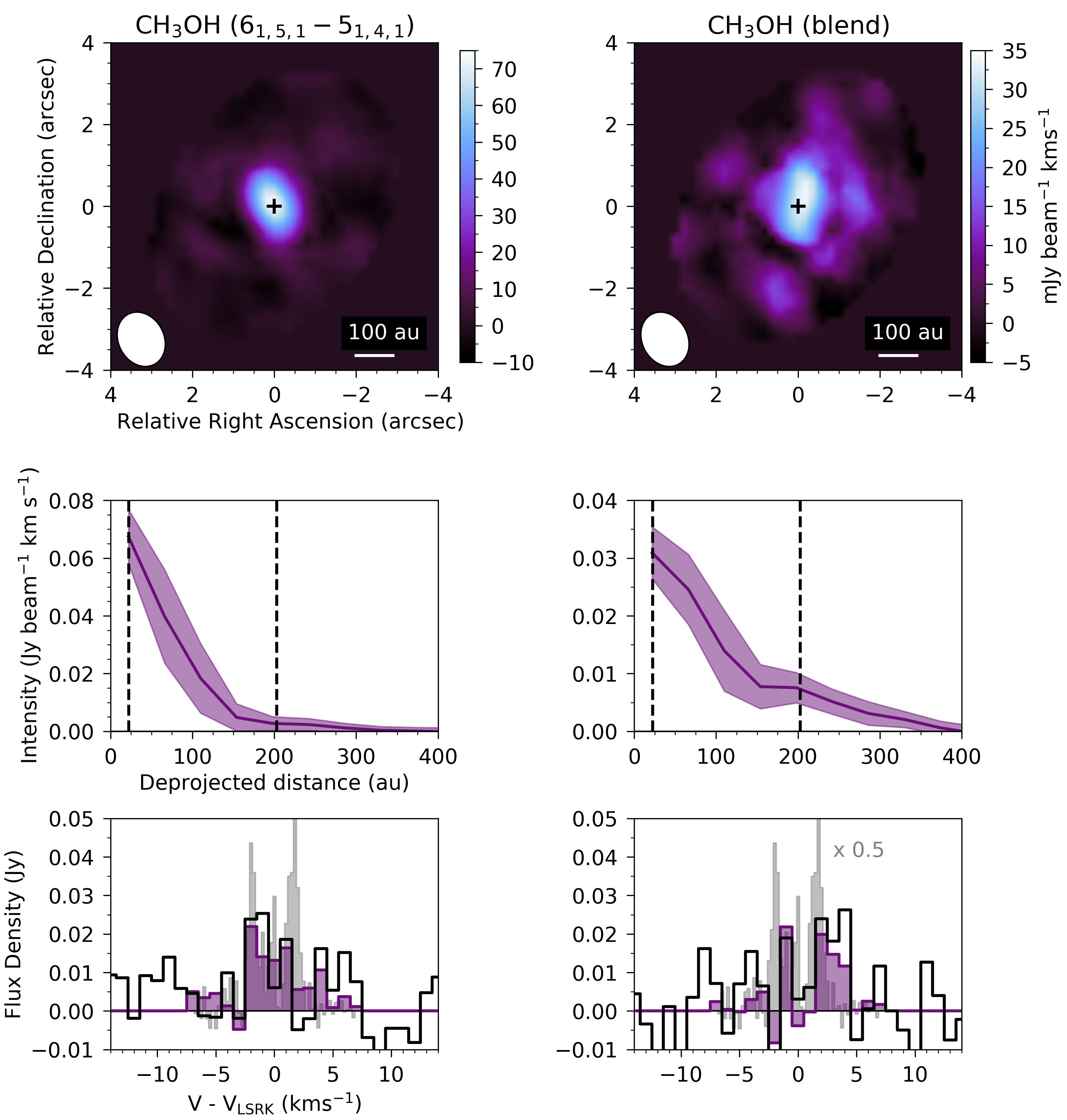}
\caption{The detected \ce{CH3OH} emission lines in the HD~100546 disk. 
Top: the integrated intensity maps made with \AB{a Keplerian mask} where the detections have a 10$\sigma$ (left) and 5$\sigma$ (right) significance where $\sigma$ is the rms noise in un--masked integrated intensity maps \AB{(7.0~mJy beam$^{-1}$ km~s$^{-1}$)}. 
Middle: the deprojected and azimuthally averaged radial profiles of the above intensity maps where the dashed lines denote the centers of the two millimeter dust rings.
Bottom: the disk--integrated spectra from  a 400~au Keplerian (purple) mask and an elliptical (black) mask. \AB{The grey shaded region is the \ce{p-H_2CO}  \ce{4_{2,3}} - \ce{3_{2,2}} transition where the flux is multiplied by 0.5 to allow for a qualitative comparison between the two species.} "Blend" refers to two \ce{CH_3OH} lines that blended. \AB{These are the \ce{6_{2,5,1}} - \ce{5_{2,4,1}} and \ce{6_{2,4,0}} - \ce{5_{2,3,0}} transitions, see Table~\ref{table2} in Methods for further details.}}
\label{fig:CH3OH_data}
\end{figure*}

The integrated intensity maps and radial profiles for the \ce{CH3OH} lines detected in the HD~100546 disk are presented in Figure~\ref{fig:CH3OH_data}. 
Most of the \ce{CH3OH} emission is spatially compact, i.e., it lies within the observering beam of 1\farcs40$\times$1\farcs11, and therefore originates from the inner $\leq 60$~au (1/2 of the beam minor axis) of the disk. For reference, Pluto's orbit ranges from $\approx 30$ to $\approx 50$~au from the Sun. There is also some diffuse emission in the outer disk hinting at a ring of \ce{CH3OH} coincident with an outer millimeter dust ring at $\approx~200$~au \citep{2014ApJ...791L...6W}.
Formaldehyde, \ce{H2CO}, which is chemically related to \ce{CH3OH} is also detected in our data (see Figure~\ref{fig:H2CO_data} in Methods). 
The \ce{H2CO} emission shows a centrally peaked component of emission and a well detected outer ring at $\approx~200$~au. 
Also shown in Figure~\ref{fig:CH3OH_data} are the spectra extracted using both a mask that follows the expected Keplerian pattern of emission with a radius of 400 au and an elliptical mask. 
The total flux for each line was calculated using the Keplerian masked spectra (see Table~\ref{table2} in Methods) and these values are used in the subsequent analysis. 

As a first estimate of the amount of gas--phase \ce{CH3OH} present in HD~100546, the disk--averaged column density of \ce{CH3OH} was derived from the single unblended transition using the same excitation temperature as calculated from the multiple \ce{H2CO} lines (see Methods for full details).
This resulted in a value of $7.1\times10^{12}~\mathrm{cm^{-2}}$ which corresponds to a gas--phase methanol mass of $6.9 \times 10^{17}$~kg within a radius of 50~au.  This is around 1,200 times more than the total estimated biomass on Earth \citep{Bar-On6506}.

\begin{figure*}
    \centering
    \includegraphics[width=0.95\hsize]{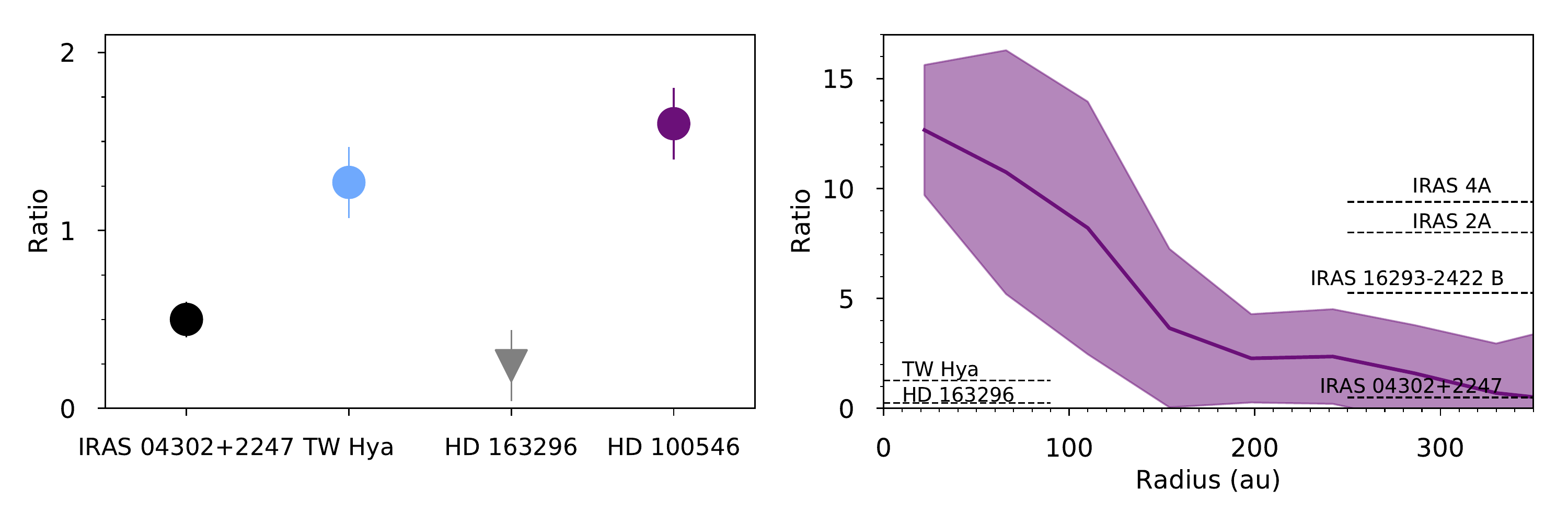}
    \caption{A comparison of the HD~100546 \ce{CH3OH} to \ce{H2CO} ratio to other sources. Left: Disk--averaged \ce{CH3OH} to \ce{H2CO} column density ratio for four disks. The bars show the error on the ratio and the arrow denotes an upper--limit.
    Right: The \ce{CH3OH} to \ce{H2CO} ratio as a function of radius for the HD~100546 disk. 
The shaded region highlights the error propagated from the radial profiles shown in Figure~\ref{fig:CH3OH_data} and Figure~\ref{fig:H2CO_data}. 
The dashed lines mark the average ratios for other objects (references in the text).}
    \label{fig:CH3OH-H2CO_ratio}
\end{figure*}

The degree of chemical processing of organic ices that may have happened in the protoplanetary disk during disk formation and evolution can be estimated from these data. Chemical models show that the abundance ratio of \ce{CH3OH} to \ce{H2CO} is a useful diagnostic of such processing, showing an increase in this ratio with time \citep{2020A&A...634A..52S}. This is because \ce{H2CO} is also formed via the hydrogenation of CO ice, en route to the formation of \ce{CH3OH}. All of the detected lines of \ce{CH3OH} and \ce{H2CO} are optically thin and so the column density ratio of \ce{CH3OH} to \ce{H2CO} can be determined (see Methods for full details). 
If the inner disk emission is more compact than the observing beam the observations may suffer from beam dilution 
and this would result in higher column densities. 
This disk--averaged ratio is shown in Figure~\ref{fig:CH3OH-H2CO_ratio} alongside the values reported for the disks around TW~Hya (a Sun like star), HD~163296 (a Herbig Ae star) and \AB{IRAS 04302+2247 (a young, $<$1~Myr Sun like star)} \citep{2016ApJ...823L..10W,2019A&A...623A.124C, 2020A&A...642L...7P}. 
The \ce{CH3OH} to \ce{H2CO} abundance ratio in the HD~100546 disk is consistent with or up to 80\%
higher than the value reported for the TW~Hya disk, at least $7$~times larger than the upper limit in HD~163296, \AB{and distinctly higher than in IRAS~04302$+$2247.}

This same column density calculation strategy was applied to the radial emission profile of the single unblended \ce{CH3OH} emission line to calculate the column density of \ce{CH3OH} as a function of radius. 
The radial \ce{CH3OH} to \ce{H2CO} ratio is also shown in Figure~\ref{fig:CH3OH-H2CO_ratio}, alongside the reported disk averaged values for the \AB{IRAS 04302+2247}, TW~Hya, HD~163296 disks. 
In the outer disk HD~100546 has a TW~Hya--like ratio showing that the species potentially have the same chemical origin in both disks. 
In the inner disk the ratio is much higher, and is more similar to the values \AB{inferred} in various young stellar objects \citep{2007A&A...465..913B, 2013A&A...554A.100I, 2018A&A...620A.170J}. 
This highlights HD~100546 as host to the most chemically complex 
protoplanetary disk in oxygen--bearing COMs discovered to date.

\begin{figure}
    \centering
    \includegraphics[width=0.7\hsize]{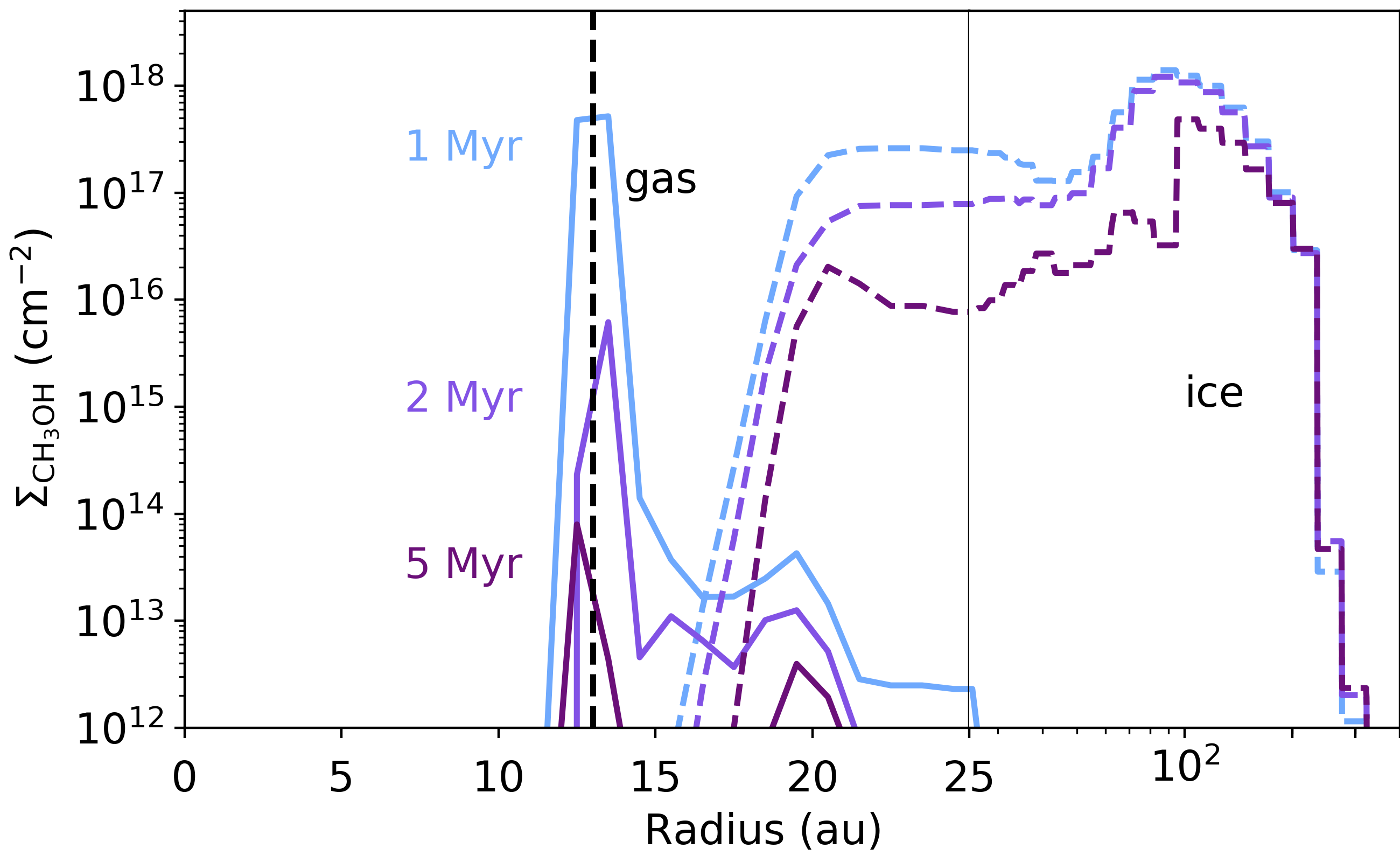}
    \caption{Model gas (solid line) and ice (dashed line) phase \ce{CH3OH} in the HD~100546 disk model as a function of radius and time. The black dashed line is the edge of the dust cavity. The solid black line marks where the x-axis scale changes from linear to log.}
    \label{fig:CH3OH_model-rad}
\end{figure}

Chemical models of protoplanetary disks predict a significantly larger mass reservoir of methanol ice than methanol gas in the outer disk, of the order of $10^{6}$ times more massive \citep{2014A&A...563A..33W}. 
To investigate the chemical origin of gas--phase \ce{CH3OH} in the HD~100546 disk we ran a gas--grain chemical model using a physical structure specific to HD~100546 with dark cloud initial molecular ice and gas phase abundances (see Methods for full details). This ice is initially \ce{CH_3OH} rich. 
We find significant abundances of gas--phase \ce{CH3OH} only in the inner disk ($< 50$~au) located at the edge of the known dust cavity ($\approx$~13~au). Here the dust is warm enough to thermally sublimate the ices formed in the cold phase \cite{2011A&A...531A..93M}. 
The radial distribution is consistent with the observations (see Figure~\ref{fig:CH3OH_data}).  
\ce{CH3OH} is destroyed via reactions with cations (e.g. \ce{H3O+})
with time and has not reached steady state by 5~Myr.   
This is because \ce{CH3OH} cannot reform efficiently in the gas phase as also shown in previous work \cite[][]{2018A&A...612A..88L}. 
The average column density in the model at 5~Myrs (the approximate age of the source) is $\approx~100\times$ less than the column derived from the observations within 50~au ($2.0\times10^{14}$~cm$^{-2}$, see Methods Figure 7). However, the column at earlier times better matches that derived from the observations. At 2~Myrs it is only $\approx$2 times less than observed. 
This indicates potentially an earlier phase of chemical and physical evolution that we are not accounting for in our static model. It has been shown that dust dynamics 
and grain growth will also have an effect on the gas and ice chemistry by continuously enriching the inner disk with \ce{CH_3OH}--rich ices  \cite{2017MNRAS.469.3994B}. 

To further investigate the formation and destruction pathways of \ce{CH3OH} in the disk we ran models with initial atomic abundances, mimicking total reset of chemistry in the disk due to e.g., shocks, in order to test if the \ce{CH3OH} is able to form in situ (see Methods for further details). We find that under these atomic initial conditions \ce{CH3OH} cannot form efficiently within 1~Myr and is $\approx10^5\times$ lower than in our inherited model at the same timestep. 
In the inner disk the inherited gas phase \ce{CH3OH} can be sustained for 1~Myr via the same gas phase reactions that are important in hot core chemistry  \citep{2016ApJ...821...46T}. 
In the outer disk no significant \ce{CH3OH} is formed 
and over time the \ce{CH_3OH} ice abundance continually decreases.
This is primarily because the dust temperature is $> 20$~K and therefore there is little CO ice present for the hydrogenation reactions to proceed to replenish CH3OH ice lost through non-thermal desorption \citep{2002ApJ...571L.173W}. 
 
\subsection*{Discussion}

Disks around Herbig Ae/Be stars have been thought to be COMs poor when compared to their T Tauri counterparts due to their warmer midplane temperatures and higher levels of UV irradiation. Chemical models support this showing significantly different CO ice reservoirs between T Tauri and Herbig Ae/Be disks with very little CO freezeout in the latter \citep{2018A&A...616A..19A, 2018A&A...618A.182B}. 
This is also consistent with \cite{2020ApJ...890..142P} who show on average, from the current source sample, that there is less \ce{H_2CO} observed in Herbig Ae/Be disks than T Tauri disks. 
In the T--Tauri source TW Hya the gas phase \ce{CH3OH} traces the CO ice reservoir whereas the disk around HD~100546 is too warm to host a significant reservoir of CO ice \cite{2016A&A...592A..83K}. The detection of \ce{CH_3OH} in the HD~100546 disk is therefore surprising, and we seek to understand its origin.

\begin{figure}[h]
    \centering
    \includegraphics[width=0.9\hsize]{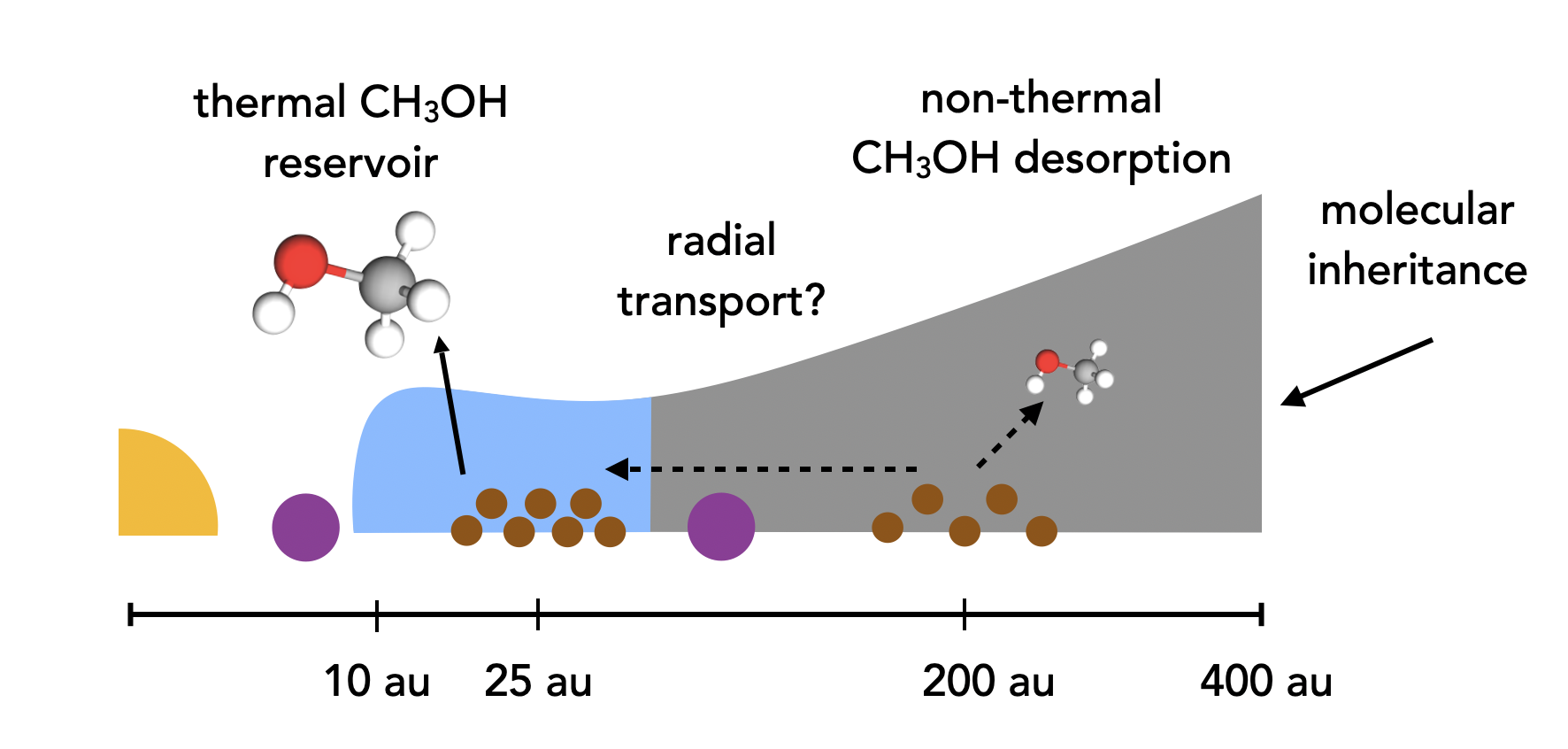}
    \caption{Cartoon of the HD~100546 disk structure showing the regions where \ce{CH_3OH} is detected in the disk and the different physical and chemical mechanisms that we propose are taking place.}
    \label{fig:disk_cartoon}
\end{figure}

In protoplanetary disks, \ce{CH3OH} can only be observed once released from the ice into the gas phase, either via thermal sublimation (desorption) in hot regions ($\gtrsim 100$~K) or via non--thermal desorption triggered by, e.g., ultra--violet photons, in cooler regions ($\lesssim 100$~K). 
In the TW~Hya disk the \ce{CH_3OH} is distributed in a ring where the outer radius is at the edge of the millimetre dust disk and the inner edge at the CO snowline \citep{2013Sci...341..630Q, 2016ApJ...823L..10W}.
At these low temperatures non--thermal processes are most likely responsible for the occurrence of the gas phase \ce{CH_3OH}. 
In the TW Hya disk the observed abundance is much lower than initially expected by chemical models \citep{2014A&A...563A..33W}. This is because the outer disk is too cold for thermal desorption and the \ce{CH_3OH} primarily fragments during the process of non--thermal desorption \citep{2018IAUS..332..395W}. 
In comparison, the bulk of the \ce{CH3OH} emission in the HD~100546 disk originates from the warmer inner disk (See Figure~\ref{fig:disk_cartoon}).
Hence it is likely a thermally desorbed reservoir of \ce{CH_3OH} that is originating from the edge of the dust cavity which is exposed to radiation from the central star.
Direct evidence for the high dust temperature at the cavity edge for the HD~100546 disk comes from the detection of crystalline silicates
\cite{2011A&A...531A..93M}.
The unique chemistry of transition disks has been predicted from models \citep{2011ApJ...743L...2C} but this is the first evidence for warm rich complex organic chemistry due to thermal sublimation in a Class II disk. 
Both TW~Hya and HD~163296 do not have large ($>5$~au) inner cavities therefore, in both cases, the non--detections of an inner warm \ce{CH3OH} component like we see in HD~100546 may be attributed to optically thick dust at millimeter wavelengths. 
Any gas--phase \ce{CH_3OH} in the outer HD~100546 disk likely has the same chemical origin as in TW~Hya, i.e., non-thermal desorption from the ice coated dust grains. 

\ce{CH3OH} has gone undetected towards both Class I and Class II disks where \ce{H2CO} is detected robustly, \AB{aside from TW~Hya and IRAS 04302$+$2247} \citep{2016ApJ...823L..10W,2020A&A...642L...7P}.
The low \ce{CH3OH}/\ce{H2CO} ratios observed in these Class I/II sources have been interpreted as evidence for chemical processing of material as it is accreted onto the disk \citep{2020arXiv200812648P}, \AB{but temperature can also be responsible for lack of emission \citep{2020ApJ...901..166V}}. 
Our detection of \ce{CH3OH} in the HD~100546 disk shows that complete chemical reset/reprocessing of ice during disk formation may not occur. Due to the nature of the HD~100546 disk, one with a central cavity, we have been able to detect a thermally desorbed reservoir of \ce{CH3OH}. This means that our gas phase abundances should be more representative of the total \ce{CH3OH} ice in the system. \AB{The lack of a similar warm \ce{CH_3OH} reservoir in the inner disk of other sources is likely due to the high optical depth of the dust which masks any thermally desorbed \ce{CH_3OH} emission. Their current upper--limits on the \ce{CH3OH} abundances} are likely probing the cold outer disk where most of the \ce{CH3OH} is on the icy grains. \AB{We also cover \ce{CH_3OH} transitions with higher upper energy levels and therefore lines that are more sensitive to warm gas.}

We use models to investigate the chemical origin of the \ce{CH_3OH} in the HD~100546 disk and, in particular investigate if \ce{CH_3OH} can be formed in--situ. 
We find that with the current observationally constrained HD~100546 disk model, it is not possible for methanol to be synthesised in non--negligible abundances anywhere in the disk when beginning with atomic initial conditions (chemical reset). This means that the disk either inherited 
\ce{CH3OH} rich ices from an earlier dark cloud phase 
or, earlier in the evolution of the disk there was a substantial cold ($< 20$~K) reservoir in the outer disk which would allow for in--situ formation. 
However, recent studies of the temperature structure of Class I disks shows that young disks are warm and thus likely cool with age \citep[e.g.,][]{2020ApJ...901..166V}.
Additionally, the bulk volatile carbon abundance in the disk is consistent with an ISM abundance \citep{2016A&A...592A..83K}. This means that 
there has been very little in--situ CO freeze--out in this disk unlike in other sources, e.g. HD~163296 and TW Hya \citep{2018A&A...618A.182B,2019ApJ...883...98Z}. 
All of the above is very strong evidence for interstellar inheritance of \ce{CH3OH} rich ice in a disk. This is the first time such a clear argument can be made as, for example, the observations of \ce{CH_3CN} in TW Hya can be explained by in--situ chemistry \citep{2018ApJ...859..131L}. 
Additionally, our models show that the \ce{CH3OH} column density does not reach steady state in the inner disk and declines steadily after 1~Myr. Therefore, it is possible that the inner disk is being replenished with \ce{CH3OH} ice from the outer disk. This could be achieved via radial transport \citep[e.g.][]{2019MNRAS.487.3998B} for which there is evidence that this is ongoing in this disk \citep{2014ApJ...791L...6W}. 



Disks around Herbig~Ae/Be stars often present numerous direct and indirect signatures of forming gas giant planets \citep[e.g.][]{2018ApJ...860L..12T, 2019NatAs...3.1109P}.
We have shown that these disks can host a substantial chemically complex reservoir that is inherited from an earlier cold phase.
Therefore, the forming cores of planets and moons in these systems will be able to accrete icy, organic--rich material
and, the building blocks of pre--biotic molecules are present at the epoch of planet formation

\newpage

\section*{Acknowledgements}

A.S.B. acknowledges the studentship funded by the
Science and Technology Facilities Council of the United
Kingdom (STFC).
C.W.~acknowledges financial support from the University of Leeds, the
Science and Technology Facilities Council, and UK Research and
Innovation (grant numbers ST/R000549/1, ST/T000287/1, and MR/T040726/1).
J.D.I. acknowledges support from the Science and Technology Facilities Council of the United Kingdom (STFC) under ST/T000287/1.
M.K. was supported by the University of Tartu ASTRA project 2014-2020.4.01.16-0029 KOMEET, financed by the EU European Regional Development Fund.





\newpage

\section*{Methods and Supplementary Data}

\subsection*{Data Reduction}

HD~100546 was observed with ALMA at Band 7 during Cycle 7 (2019.1.00193.S, full details are shown in Table~\ref{table1}). The data were processed using the Common Astronomy Software Package version 5.6.1 \citep[CASA;][]{2007ASPC..376..127M}. The source was observed across 11 spectral windows. During the inspection of the data it was noted that an additional antenna needed to be flagged in the calibration script which used the bandwidth--switching method. 
Antenna DA46 was flagged manually using CASA task \texttt{flagdata} and the calibration script was re--run. The bandpass and flux calibrators are the same, J1107--4449. 
The line data were then self--calibrated by splitting out the continuum--only channels from the individual spectral windows. Two rounds of phase calibration were applied, and then one of amplitude calibration. 
These calibration tables were then applied to the full line spectral windows. The data were then phase centred using \texttt{uvmodelfit} and \texttt{fixvis}. 
The continuum subtraction was performed using \texttt{uvcontsub} with fit order of 0. 
The eight spectral lines analysed in this work are covered in 7 spectral windows and are listed in Table~\ref{table2} along with their transition information. 
The native spectral resolution of these spectral windows is 122.070~kHz with a total of 480 channels and after Hanning smoothing this is equivalent to a velocity resolution of 0.25~$\mathrm{km~s^{-1}}$. 
The two \ce{o-H2CO} lines are each covered in their own spectral window but are blended since their rest frequencies are only $\approx~$4.0~$\mathrm{km~s^{-1}}$ apart. \AB{This means that the emission from the two transitions is overlapping in the line wings between the two reset frequencies.} 
These measurements sets were therefore combined using \texttt{concat}.
The \ce{CH3OH(E)} \ce{6_{1,5,1}} - \ce{5_{1,4,1}} and \ce{6_{2,4,0}} - \ce{5_{2,3,0}} transitions are covered in the same spectral window and are also blended with their rest frequencies  $\approx~$0.4~$\mathrm{km~s^{-1}}$ apart.
\AB{This means that the frequency resolution of the observations is not sufficient to distinguish between the emission attributed to the individual transitions.}

\begin{figure}
    \centering
    \includegraphics[width=0.9\hsize]{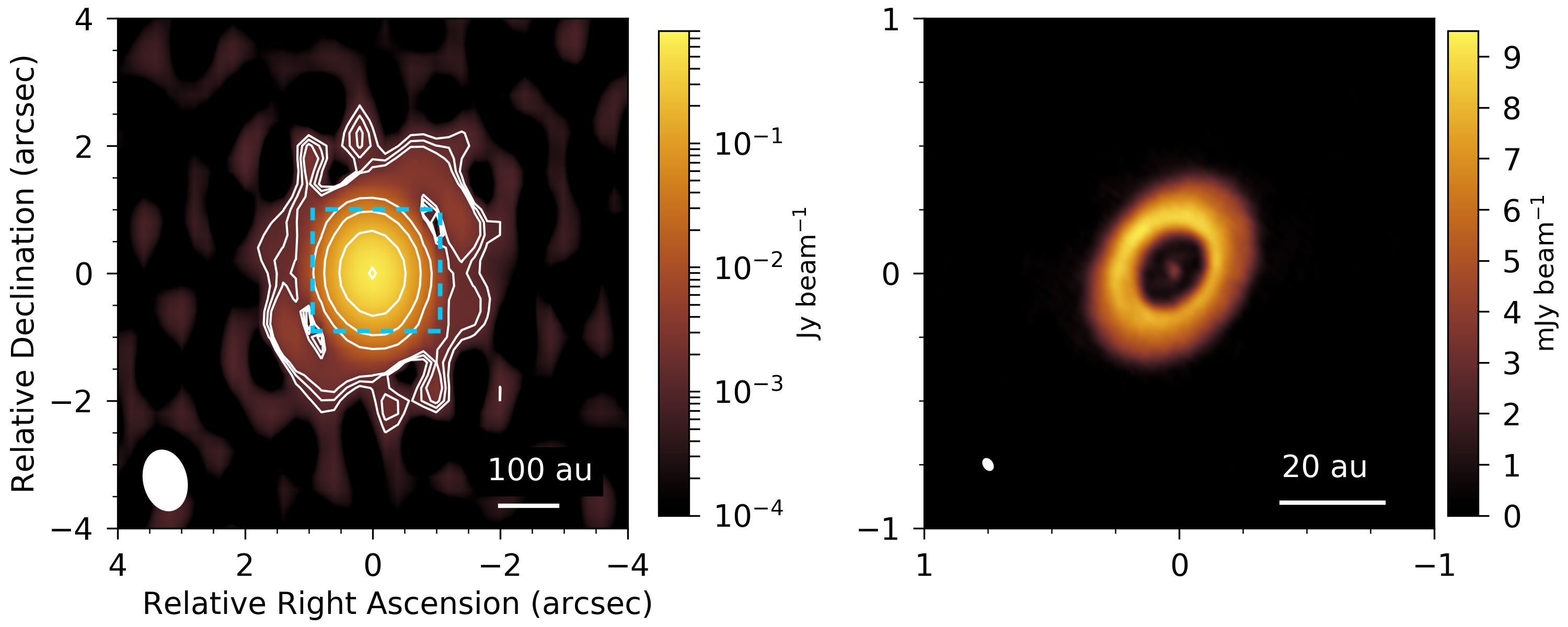}
    \caption{Images of the continuum emission at 290~GHz (left; this work) and at 345~GHz (right; \cite{2019ApJ...871...48P}). The right--hand image is a 2\farcs0 square zoom of the inner disk at higher angular resolution (this region is highlighted by the dashed box on the right panel) revealing the resolved inner disk cavity with a radius of 13~au where the bulk of the gas phase \ce{CH_3OH} is located. The white contours on the left image are at [5,7,9,100,300,1000,3000,6000]$\times~\sigma$ where $\sigma$ is 0.22~mJy$\mathrm{beam^{-1}}$ }
    \label{fig:continuum}
\end{figure}

\begin{figure*}
    \centering
    \includegraphics[width=\hsize, trim=5cm 0cm 3cm 0cm, clip]{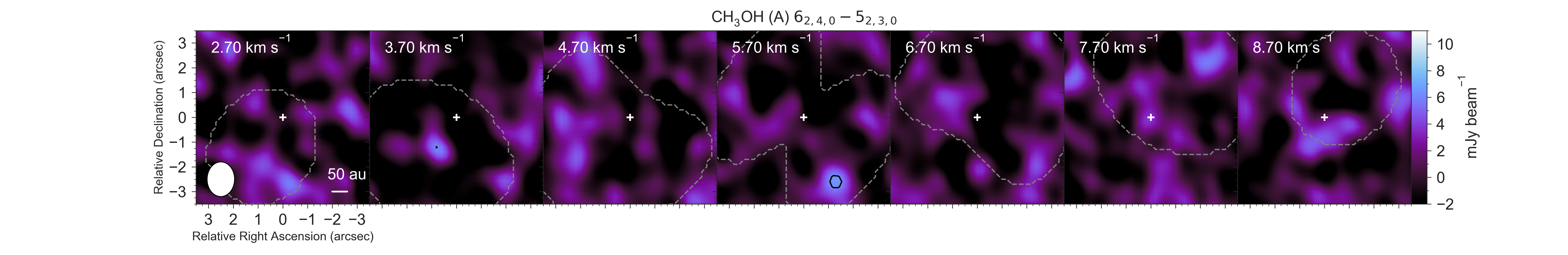}
    \includegraphics[width=\hsize, trim=5cm 0cm 3cm 0cm, clip]{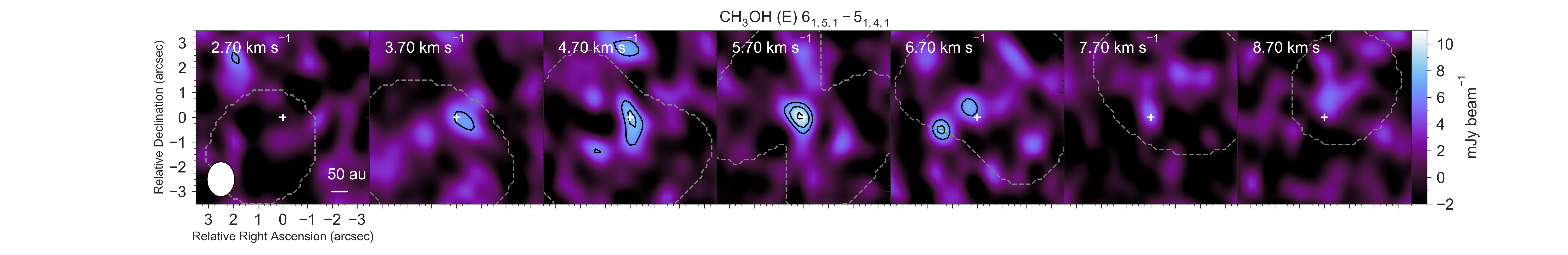}
    \includegraphics[width=\hsize, trim=5cm 0cm 3cm 0cm, clip]{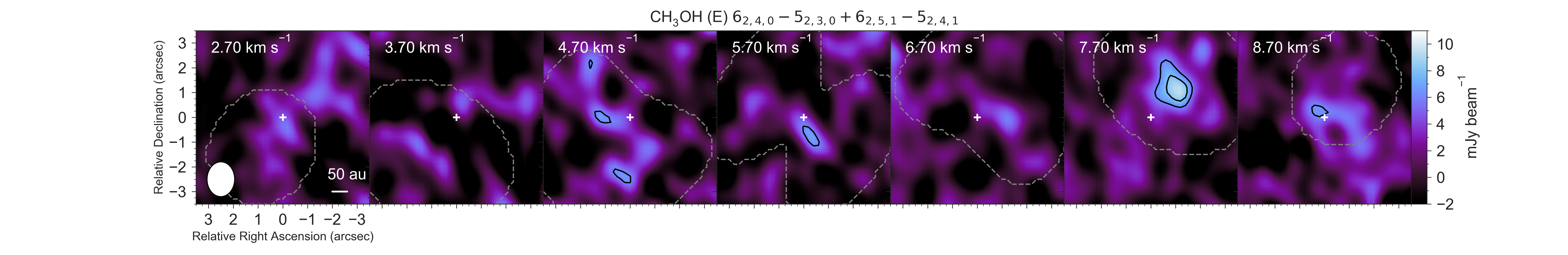}
    \caption{\AB{\ce{CH_3OH} channel maps at a velocity resolution of 1~km~s$^{-1}$. The black contours mark the [3,4,5]$\times~\sigma$ levels where $\sigma$ is as stated in Table~\ref{table2}. The Keplerian mask used in tCLEAN and to make the integrated intensity maps is marked by the grey dashed line.}}
    \label{channelmaps}
\end{figure*}

All of the molecular lines were imaged with \texttt{tCLEAN} using a 400~au Keplerian mask (approximately the same size as the \ce{^{12}CO} molecular disk) and natural weighting. The resulting rms, peak line flux, and integrated flux for each line are listed in Table~2 along with the respective beam sizes. The \ce{p-H2CO} and \ce{o-H2CO} lines are detected clearly at the native spectral resolution of the data whereas the \ce{CH3OH(E)} data needed to be re--binned to $\mathrm{1~km s^{-1}}$ in order to make a clear ($\geq5\sigma$) detection. The \ce{CH3OH(A)} \ce{6_{2,4,0}} - \ce{5_{2,3,0}} transition was not detected at $\mathrm{1~km s^{-1}}$ nor when using a matched filter analysis \citep{2018AJ....155..182L}. 
This line has a slightly higher upper energy level and lower transition probability than the detected \ce{CH3OH} lines (see numbers for all transitions covered are in Table~\ref{table2}), and because the signal--to--noise for the detected lines is low, the non--detection of this line is not unreasonable. \AB{The channel maps of all \ce{CH_3OH} lines with the Keplerian masks overlaid are shown in Figure~\ref{channelmaps}.}

The \ce{CH3OH} integrated intensity maps shown in Figure~\ref{fig:CH3OH_data} were generated with Keplerian masks. These were then deprojected and azimuthally averaged assuming a disk inclination of 44\degree~and position angle of 146\degree~\citep{2014ApJ...791L...6W}. 
The resulting radial profiles are presented in Figure~\ref{fig:CH3OH_data} where the associated errors are the standard deviation of the pixel values in each bin divided by the square root of the number of beams per annulus \citep[e.g.][]{2019A&A...629A..75B}. 
The equivalent data products for \ce{H2CO} are presented in Figure~\ref{fig:H2CO_data}. 
We also present an image of the continuum emission at 290~GHz from these data imaged using super--uniform weighing resulting
in a beam size 0\farcs95$\times$0\farcs66 (12\degree)
of with an rms of 0.22~mJy~$\mathrm{beam^{-1}}$ and a peak 
emission of 0.68~Jy~$\mathrm{beam^{-1}}$
(see Figure~\ref{fig:continuum}, left--hand panel).  
We also show in Figure~\ref{fig:continuum} a higher angular resolution image with a beam size of 0\farcs045$\times$0\farcs031 (33.9\degree) that well resolves the inner 13~au mm dust cavity \citep{2019ApJ...871...48P}.

The total flux for each transition was extracted using the CASA task, \texttt{SpecFlux}, and the Keplerian masks. 
Due to blending, the \ce{o-H_2CO} lines were separated via spectral shifting and stacking \citep[e.g.][]{2016ApJ...832..204Y,2019ApJ...876L..13S}.
In this method the spectra extracted from each pixel were shifted according to its Keplerian velocity projected to the line of sight. The individual spectra are then added together. 
The stacked spectra for \ce{o-H2CO} extracted from within the same area as the elliptical mask spectra are shown in Figure~\ref{fig:H2CO_data}.
The total flux for each of the \ce{o-H_2CO} lines is determined by integrating under each of the peaks in the stacked spectra. 
The measured fluxes for all of the lines are listed in Table~\ref{table2} along with the peak line flux per channel and the rms for each line. 

\begin{table*}
\centering
\begin{threeparttable}
\caption{ALMA Observations of the HD~100546 disk}
\label{table1}
\begin{tabular}{ccccc}
\hline\hline
\multicolumn{1}{l}{Observations}             & \multicolumn{4}{c}{ALMA Cycle 7 (2019.1.00193.S, PI Booth)} \\
\multicolumn{1}{l}{Source}                   & \multicolumn{4}{c}{HD~100546}     	\\ 
\multicolumn{1}{l}{Date observed}            & \multicolumn{4}{c}{10th December 2019}  \\
\multicolumn{1}{l}{Baselines (m)}            & \multicolumn{4}{c}{$15.1 - 312.7$}    \\
\multicolumn{1}{l}{On source time (mins)}    & \multicolumn{4}{c}{34.5}   \\
\multicolumn{1}{l}{Number. antennae}              & \multicolumn{4}{c}{42}   \\
\multicolumn{1}{l}{Natural CLEAN beam}              & \multicolumn{4}{c}{1\farcs40$\times$1\farcs11(28\degree)}   \\
\hline
\end{tabular}
\end{threeparttable}
\end{table*}

\begin{landscape}
        \centering 
   \begin{table}[]
   
   \footnotesize
   
\begin{threeparttable}
\caption{Molecular lines and properties}
\label{table2}
\begin{tabular}{clccccccc}
\hline \hline
Molecule & Transition & Frequency & $\mathrm{E_{up}}$ & $\mathrm{E_{A}}$   & $\Delta$v      & rms                             & Peak                              & Flux \\  
         &            &  (GHz)    & (K)              & $\mathrm{(s^{-1})}$ &        (km s$^{-1}$)  & (mJy beam$^{-1}$) &   (mJy beam$^{-1}$) & (mJy km s$^{^{-1}}$) \\ 
\hline 
\ce{p-H2CO}   &   \ce{4_{2,3}} - \ce{3_{2,2}}      & 291.238 & 82.1  & $5.210 \times 10^{-4}$ &  0.25 & 3.3 & 64.7 & 246\\
\ce{p-H2CO}   &   \ce{4_{2,2}} - \ce{3_{2,1}}      & 291.948 & 82.1  & $5.249 \times 10^{-4}$      & 0.25 & 3.2 & 70.3 & 303\\
\ce{o-H2CO}$^{+}$   &   \ce{4_{3,2}} - \ce{3_{3,1}}      & 291.380 & 140.9 & $3.044 \times 10^{-4}$       & 0.25 & 2.8 & 62.9 & 79 \\
\ce{o-H2CO}$^{+}$   &   \ce{4_{3,1}} - \ce{3_{3,0}}      & 291.384 & 140.9 & $3.044 \times 10^{-4}$         & 0.25 & 2.8 & 27.7  & 95 \\ 
\hline 

\ce{CH3OH} (A)  &  \ce{6_{2,4,0}} - \ce{5_{2,3,0}}   & 290.264 & 86.5  & $9.502 \times 10^{-5}$       & 1.0& 2.0 & - &   $<$67\\
\ce{CH3OH} (E)    &  \ce{6_{1,5,1}} - \ce{5_{1,4,1}}   & 290.249 & 69.8  & $1.058 \times 10^{-4}$         & 1.0& 2.0 & 10.4 &   98 \\
\ce{CH3OH} (E)     &    \ce{6_{2,5,1}} - \ce{5_{2,4,1}} & 290.308 & 71.0  & $9.346 \times 10^{-5}$        & 1.0& 2.0 & 9.4 & 70$^*$  \\
\ce{CH3OH} (E)   &    \ce{6_{2,4,0}} - \ce{5_{2,3,0}} & 290.307 & 74.7  & $9.458 \times 10^{-5}$&  1.0& 2.0 & 9.4 & 70$^*$ \\
\hline
\end{tabular}
\begin{tablenotes}\footnotesize
\item{The values for the line frequencies, Einstein A coefficients, and upper energy levels ($E_{up}$) 
are from the Leiden Atomic and Molecular Database: \url{http://home.strw.leidenuniv.nl/~moldata/} \citep[LAMDA;][]{2005A&A...432..369S}.}
\item{\AB{$^{+}$ these lines are partially blended}}
\item{$^{*}$ total flux of the blended \ce{6_{2,5,1}} - \ce{5_{2,4,1}} and \ce{6_{2,4,0}} - \ce{5_{2,3,0}} transitions}
\end{tablenotes}
\end{threeparttable}
\end{table}  
\end{landscape}

\begin{landscape}
        \centering 
        \begin{figure}
            \centering
            \includegraphics{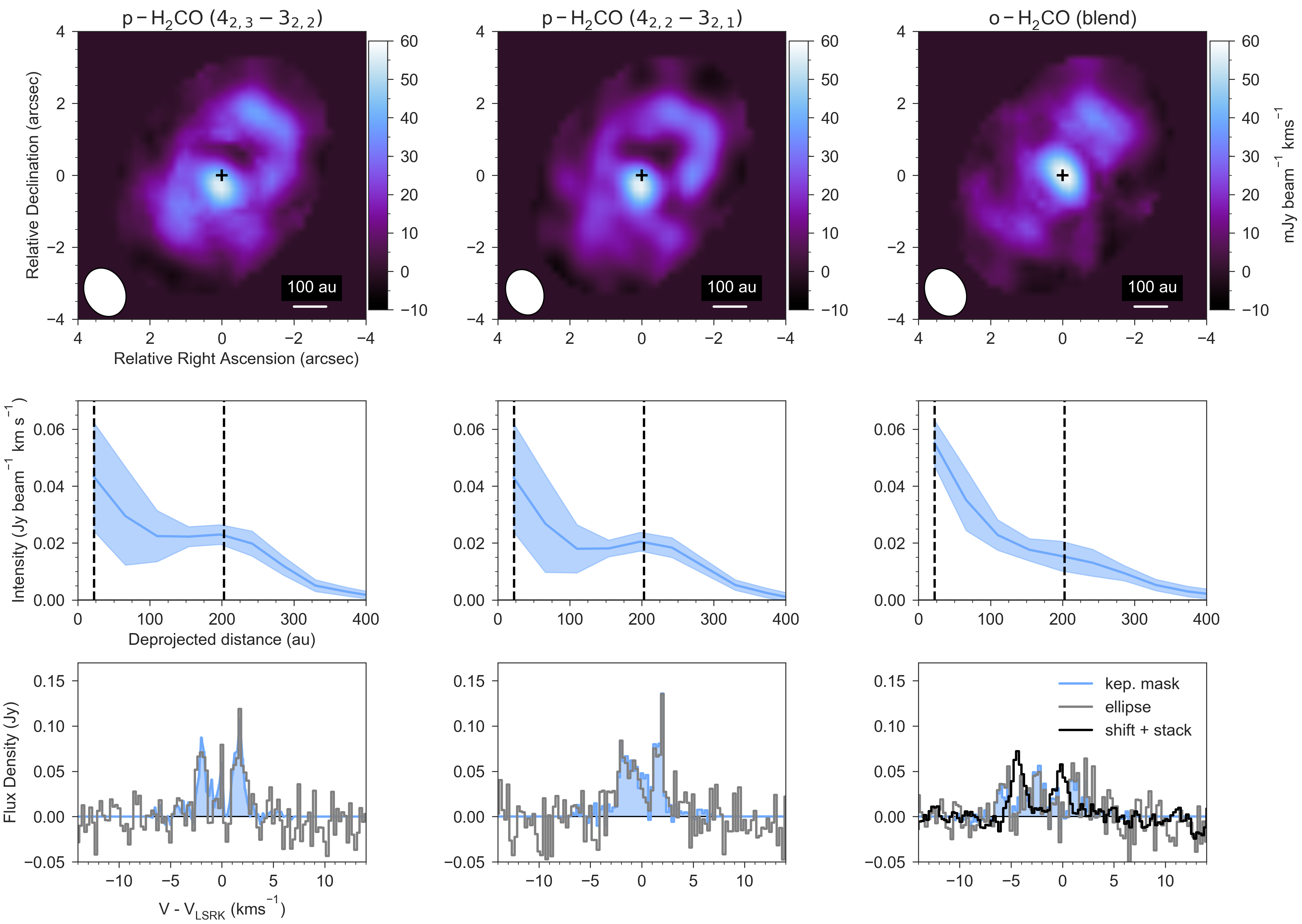}
            \caption{ The detected \ce{H_2CO} transitions in the HD 100546 disk. Top: Keplerian masked
integrated intensity maps. Middle: the deprojected and azimuthally averaged radial profiles
of the above intensity maps where the dashed lines denote the center of the two millimeter
dust rings. Bottom: integrated spectra from both a 400 au Keplerian (purple) and elliptical
(grey) mask, and velocity shifted and stacked spectra (black).
\AB{The blended \ce{o-H_2CO} transitions are the \ce{4_{3,2}} - \ce{3_{3,1}} and \ce{4_{3,1}} - \ce{3_{3,0}}}.}
            \label{fig:H2CO_data}
        \end{figure}
\end{landscape}

\subsection*{Column Density and Excitation Calculations}

\AB{As} multiple lines of \ce{H_2CO} with different upper energy levels are detected we can calculate a disk averaged rotation diagram following the methods in \citep{2018ApJ...859..131L} and \citep{2020ApJ...890..142P}.
As the lines are expected to be in the optically thin limit, the line intensity, $I_{\nu}$, is related to the column density
of the upper--state level, $N_{\mathrm{u}}^{\mathrm{thin}}$, via
\begin{equation}
I_{\nu}=\frac{A_{u l} N_{u}^{\text {thin }} h c}{4 \pi \Delta v},
\end{equation} 
where $A_{u l}$ is the Einstein A coefficient for the transition and $\Delta v$ is the line--width. 
The line intensity is given by the integrated flux $S_{\nu}$ divided by the solid angle of the beam $\Omega$. Therefore, taking this into account, Equation 1 can be rearranged to:
\begin{equation}
N_{u}^{\mathrm{thin}}=\frac{4 \pi S_{\nu} \Delta v}{A_{u l} \Omega h c}.
\end{equation} 
The column density of the upper state level can be corrected to account for the emission not being fully optically thin,
\begin{equation}
N_{u}=N_{u}^{\mathrm{thin}} \frac{\tau}{1-e^{-\tau}},
\end{equation} 
where $N_{u}$ is the corrected value and $\tau$ is the opacity at the line centre. 
This can be calculated via
\begin{equation}
\tau=\frac{A_{u l} N_{u}^{\text {}} c^{3}}{8 \pi \nu^{3} \Delta v}\left(e^{h \nu / k T_{\text {rot }}}-1\right) 
\end{equation} 
where $\nu$ is the rest frequency of the line.
The doppler line--width is given by
$$\Delta v_{l} = \sqrt{\left(\sqrt{\frac{2 k_{B} T_{e x}}{m_{X}}}\right)^{2}+\left(t_{0} \sqrt{\frac{k_{B} T_{e x}}{\mu m_{H}}}\right)^{2}},$$
where $m_X$ is the molecular mass (either for \ce{H_2CO} or \ce{CH_3OH}), $m_H$ is the mass of hydrogen, $\mu=2.37$ and 
$t_{0}$ is the turbulent component and assumed to be 0.01.
The average doppler line width between at a $T_{rot}$ of 30 and 60~K was used in the calculations (0.17 km~s$^{-1}$).
The line width / FWHM of the line is then $2 \sqrt{ln(2)} \times$ $\Delta v_{l}$. 
The total column density, $N_{tot}$, can finally be calculated from Boltzmann equation,
\begin{equation}
\frac{N_{u}^{thin}}{g_{u}}=\frac{N_{\mathrm{tot}}}{Q\left(T_{\mathrm{rot}}\right)} e^{-E_{u} / k T_{\mathrm{rot}}},
\end{equation}
where $g_u$ is the degeneracy of the corresponding upper
state level, $Q$, the partition function, $E_u$, the
upper state level energy, and  $T_{\mathrm{rot}}$, the rotational temperature.
We then create a likelihood function from Equation 5 and use emcee \citep{2013PASP..125..306F} to retrieve posterior distributions for $N^{\mathrm{total}}$, $T_{\mathrm{rot}}$, and $\tau$.

\begin{figure}
    \centering
    \includegraphics[width=0.6\hsize]{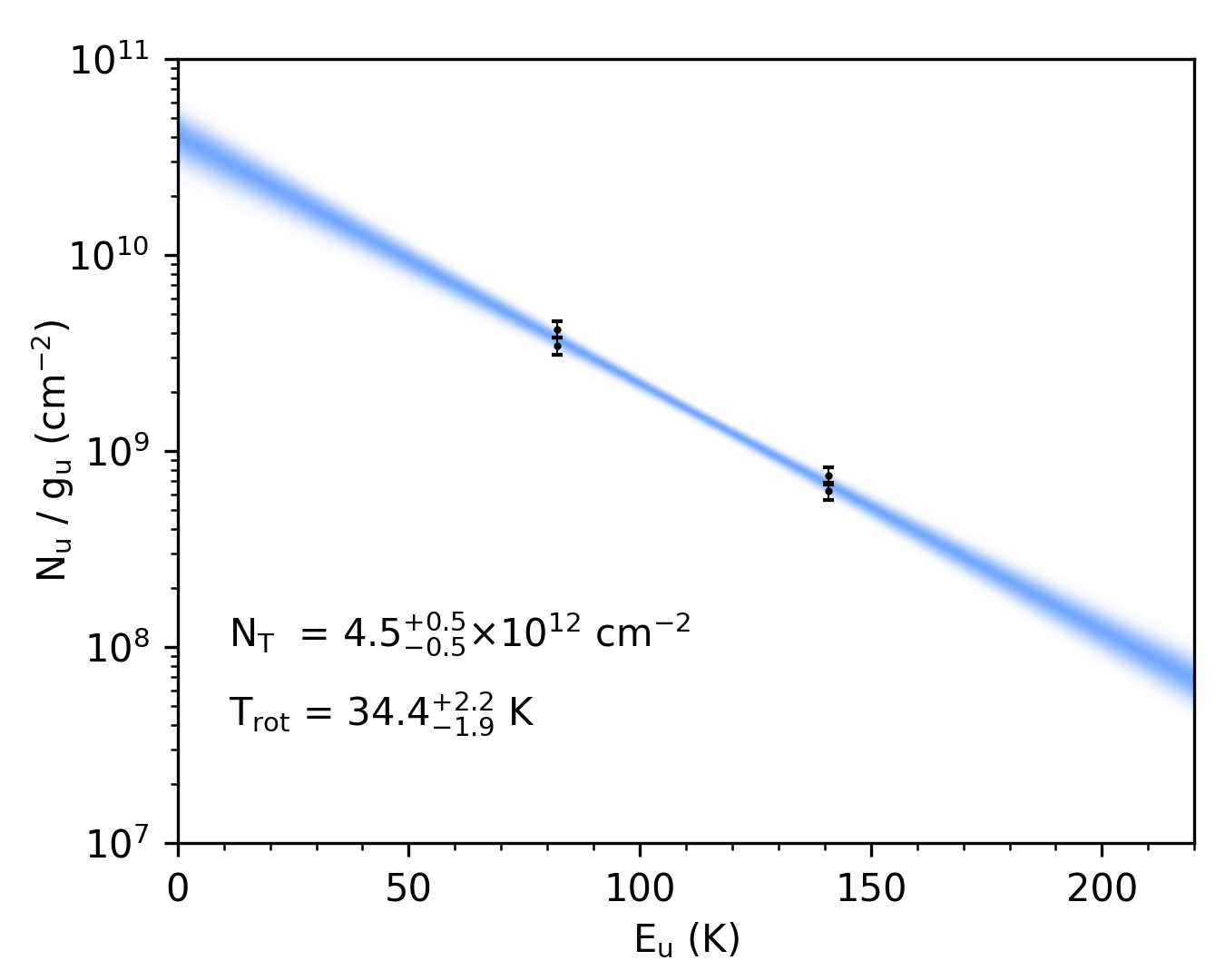}
    \caption{Disk--averaged rotation diagram for \ce{H2CO} using the flux density values listed in Table~\ref{table2} for the four detected transitions. The data points are the black markers. The coloured lines are random draws from the posterior distribution that resulted from the MCMC fitting.}
    \label{fig:H2CO_rot}
\end{figure}

First the disk integrated intensities of the \ce{H_2CO} lines were used to calculate the disk averaged \ce{H_2CO} column density and excitation temperature including a 10\% flux uncertainty. We then use this excitation temperature to calculate the \ce{CH_3OH} disk averaged column density from the \ce{6_{1,5,1}} - \ce{5_{1,4,1}} transition. This assumes that both molecules are emitting from the same vertical and radial region in the disk. 
Because all lines are found to be optically thin and have a similar emission morphology, and may have a similar chemical origin, this is a reasonable assumption.  However, given that there is evidence for radial transport in this disk, this may have redistributed methanol ice and gas through the disk relative to \ce{H2CO} gas. Higher angular resolution data would be needed to confirm this. 
The upper state degeneracy and energy, Einstein A coefficient, and frequency of
each transition and partition functions were taken from CDMS for both \ce{H_2CO} and \ce{CH_3OH} \citep{2005JMoSt.742..215M}. In this calculation we do not distinguish between the ortho and para spin states of \ce{H_2CO} and the E and A types of \ce{CH_3OH}.
The resulting disk averaged \ce{H2CO} rotation diagram is shown in Figure~\ref{fig:H2CO_rot} where the disk--averaged column
density is $4.5^{+0.6}_{-0.5}\times 10^{12}$ $\mathrm{cm^{-2}}$ and rotational temperature is $34^{+2}_{-2}$~K.
The disk--averaged optical depth of the \ce{H2CO} transitions ranged from 0.004 to 0.013.
For \ce{CH3OH} the disk--averaged column density is $7.1^{+0.7}_{-0.6}\times 10^{12}$ $\mathrm{cm^{-2}}$ and $\tau$ is 0.008.
The above error bars are propagated from a 10\% uncertainty on the flux. 
Uncertainties are the $\mathrm{16^{th}}$ and  $\mathrm{84^{th}}$ percentiles of the posterior distributions of the MCMC, corresponding to 1$\sigma$.
\AB{If the excitation temperature of the \ce{CH_3OH} is higher then the inferred column density will also be higher. For example, at 100~K the column density of \ce{CH_3OH} will increase by $\times~1.7$ and for 150~K by $\times~2.5$. We cannot constrain the \ce{CH_3OH} excitation with our current data but our inferred column densities are robust and realistic lower limits.}

\begin{figure}
    \centering
    \includegraphics[width=1.0\hsize]{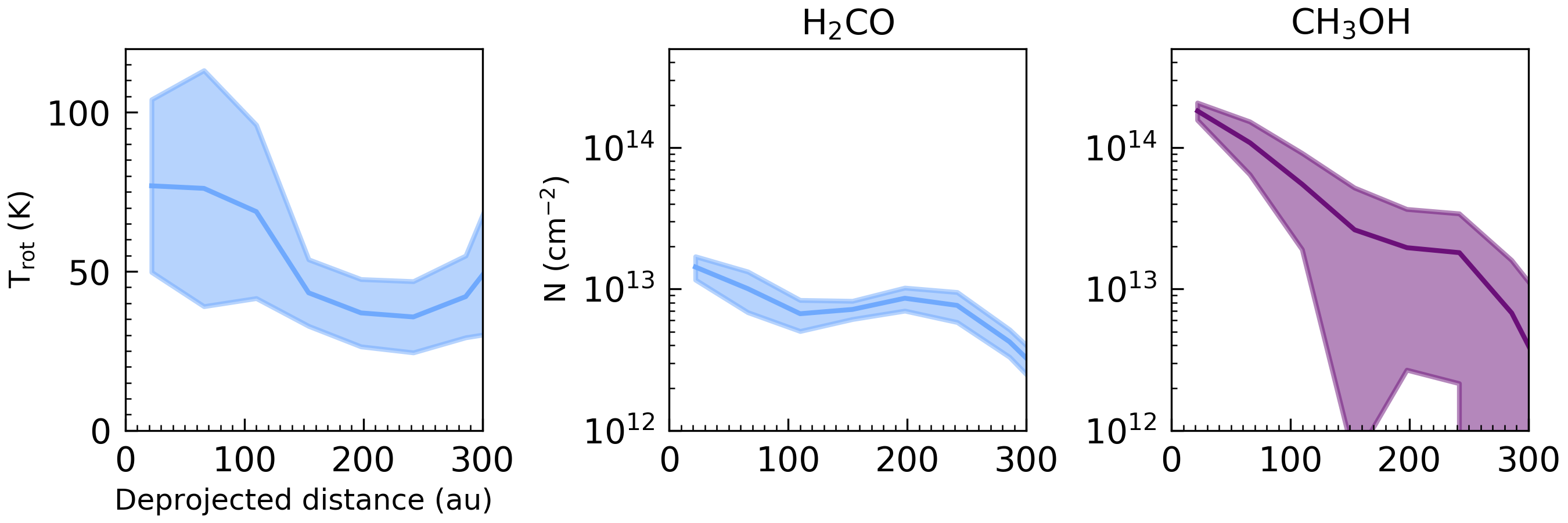}
    \caption{Radially resolved \ce{H2CO} $T_{rot}$ (left), \ce{H2CO} column density (middle) and \ce{CH3OH} column density (right). All obtained from the rotational diagram analysis using the radial emission profiles in Figures~\ref{fig:CH3OH_data} and \ref{fig:H2CO_data}.}
    \label{fig:H2CO-CH3OH_rad}
\end{figure}

Because the molecular emission is radially resolved, we can extend the above analysis to derive $N(\ce{H2CO})$,  $N(\ce{CH3OH})$ and $T_{\mathrm{rot}}$ as a function of projected radius using the radial profiles shown in Figures~\ref{fig:CH3OH_data} and \ref{fig:H2CO_data}. 
The resulting $N(\ce{H_2CO})$ and $T_{\mathrm{rot}}$ as a function of radius are shown in Figure~\ref{fig:H2CO-CH3OH_rad}. 
This radial temperature profile is then applied to the \ce{CH3OH} \ce{6_{1,5,1}} - \ce{5_{1,4,1}} radial profile and the resulting column densities are shown in Figure~\ref{fig:H2CO-CH3OH_rad}. 
Here the errors come from the errors in the radial emission profiles. 
The optical depth of the \ce{H2CO} lines reaches a maximum of 0.026 in the inner disk and a maximum of 0.25 for \ce{CH3OH}.
These radial column density profiles are then used to calculate the \ce{CH_3OH}/\ce{H_2CO} ratio presented in Figure~\ref{fig:CH3OH-H2CO_ratio}. 
\AB{Note that these are likely lower limits on the column density and line opacity in the inner disk due to beam dilution; however, the dilution factor is the same for \ce{CH_3OH} and \ce{H_2CO} if the emitting areas are the same. The column density ratio would therefore stay the same unless unless one transition becomes optically thick whilst the other does not.}

\subsection*{Chemical Modelling}

\begin{figure}
    \centering
    \includegraphics[width=1.0\hsize]{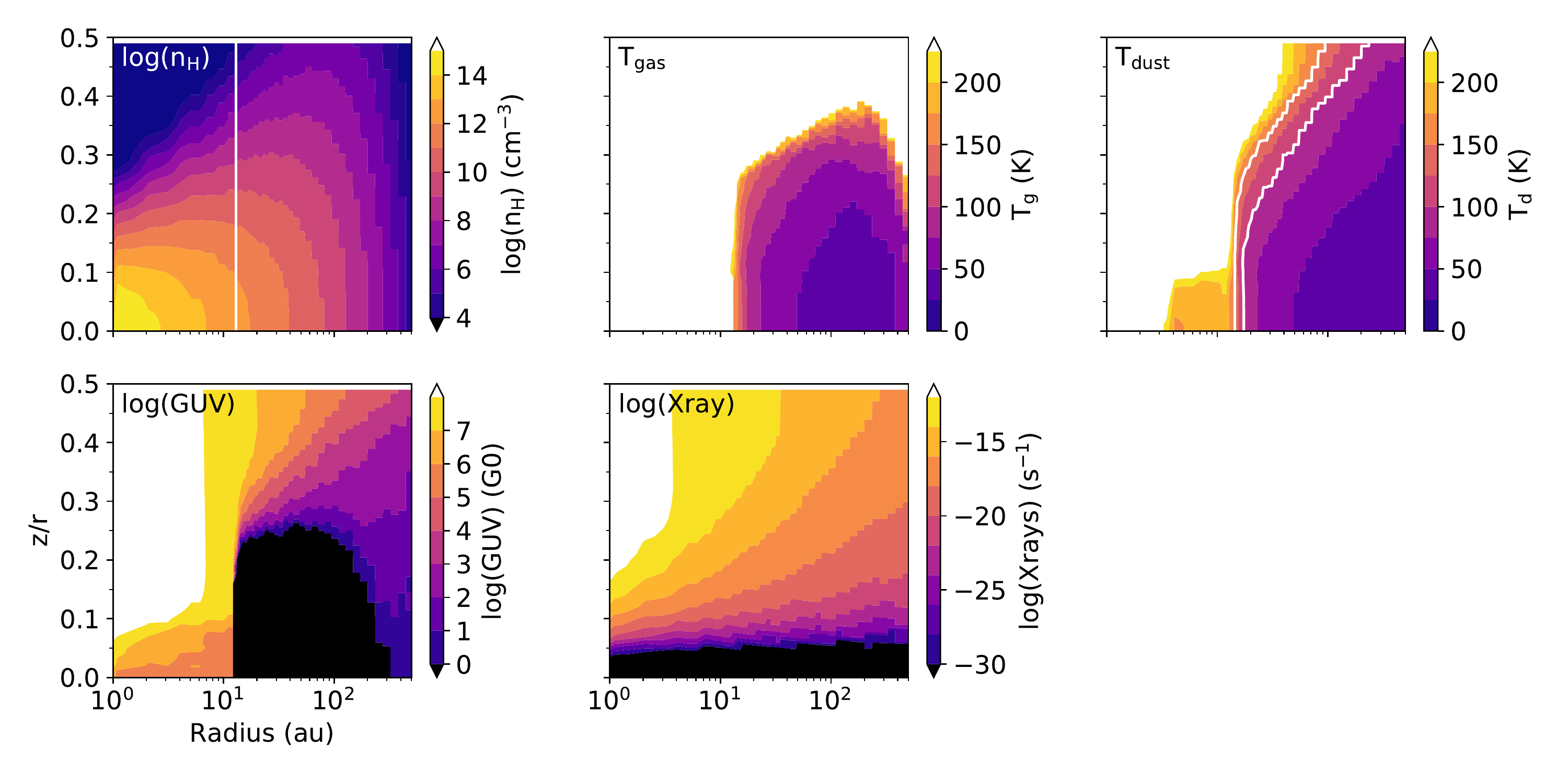}
    \caption{HD~100546 2D disk model from \cite{2016A&A...592A..83K} showing the 
   $n_\mathrm{H}$ density ($\mathrm{cm^{-3}}$), gas temperature (K), dust temperature (K), UV field (in units of the interstellar radiation field, G$_{0}$)  and X--ray ionization rate ($\mathrm{s^{-1}}$).
   The white contours denote the $T_{dust}$ of 100 and 150~K and the location of the dust cavity.}
    \label{fig:disk_model}
\end{figure}

To investigate the chemical origin of gas--phase \ce{CH3OH} in the HD~100546 disk we ran a gas--grain chemical model over a physical structure specific to the HD~100546 disk. 
The aim of this calculation is not to make a model to reproduce the observations but to explore where gas--phase \ce{CH3OH} is predicted to reside in the disk and if the predicted location and column density are similar to the observations. 
The 2D disk physical model (see Figure~\ref{fig:disk_model}) is from \cite{2016A&A...592A..83K} and this includes the $n_H$ density (number density of hydrogen nuclei), gas and dust temperature, UV radiation field and X--ray ionisation rate. 

\begin{figure}
    \centering
    \includegraphics[width=1.0\hsize]{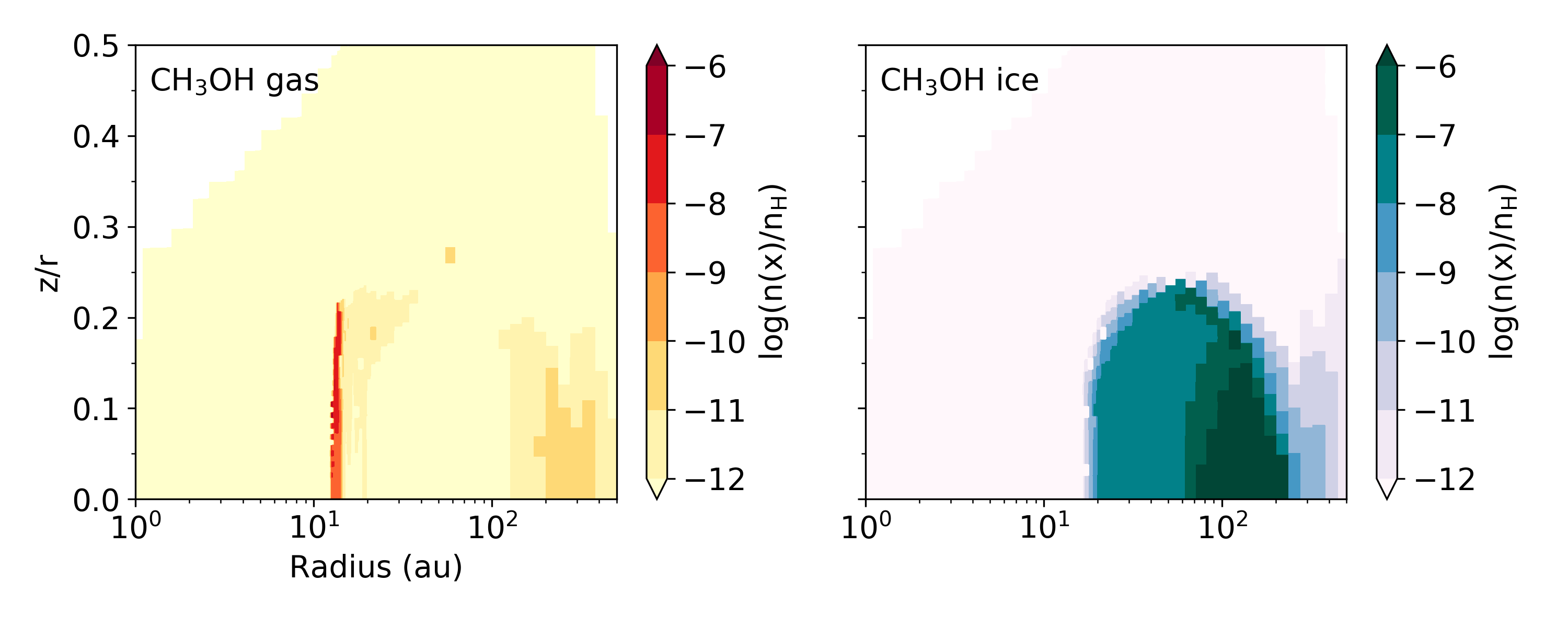}
    \caption{Fractional abundances with respect to $n_{\mathrm{H}}$ in the HD~100546 disk model at 1~Myr for \ce{CH_3OH} gas and ice assuming the inheritance scenario.}
    \label{fig:CH3OH_model-points}
\end{figure}

Following that assumed in \cite{2016A&A...592A..83K} we also assume a cosmic ray ionisation rate of $5\times10^{-17}$~s$^{-1}$. 
We couple this model with the gas--phase and gas--grain reaction network presented in \cite{2010ApJ...722.1607W, 2012ApJ...747..114W, 2013ApJ...766L..23W, 2014A&A...563A..33W, 2015A&A...582A..88W}.
The gas--phase chemistry network is based on the UMIST Database for Astrochemistry \cite[Rate 12;][]{2013A&A...550A..36M} that was first implemented into the code in \citep{2014MNRAS.445..913D, 2015MNRAS.451.3836D}. 
The accretion and desorption rates are from \cite{1982A&A...114..245T} and the grain surface reaction rates are from \cite{1992ApJS...82..167H, 2008ApJ...682..283G, 2011ApJ...735...15G}. In the modelling framework we also assume a gas--to--dust mass ratio of 100 at every point in the disk with surface reactions assuming to occur on dust grains of radius 0.1~$\mu$m.
The \ce{CH3OH} chemistry was recently updated \cite{2018IAUS..332..395W} to include the fragmentation of \ce{CH3OH} upon non--thermal photo--desorption \cite{2016ApJ...817L..12B} and the surface chemistry network for \ce{CH_3OH} was extended \cite{2016MNRAS.455.1702C}. 
This results in a total of 709 species with 9441 reactions. 
The initial molecular abundances were determined by running a single--point dark--cloud chemical model from atomic initial conditions for 1~Myr. 
The gas and dust temperature were set to 10~K, and a density of H nuclei of $\mathrm{2.0\times10^{4}~cm^{-3}}$ was assumed, and the cosmic ray ionisation rate was set to $\mathrm{10^{-17}~s^{-1}}$.
The use of these initial abundances assumes the inheritance of the ice from an earlier cold phase. In particular the initial fractional abundance of select ice species relative to n$_{\mathrm{H}}$ is as follows:
\ce{CH_3OH}~$=2.9\times10^{-8}$ $\mathrm{cm^{-3}}$,
\ce{H_2O}~$=1.5\times10^{-4}$ $\mathrm{cm^{-3}}$,
\ce{CO}~$=3.6\times10^{-5}$ $\mathrm{cm^{-3}}$,
\ce{CO_2}~$=6.7\times10^{-7}$ $\mathrm{cm^{-3}}$,
\ce{CH_4}~$=4.2\times10^{-6}$ $\mathrm{cm^{-3}}$,
\ce{NH_3}~$=1.4\times10^{-6}$ $\mathrm{cm^{-3}}$.
The chemistry was then run for each point in the disk model for 10~Myrs. 
The 2D output at 1~Myr, the time when the chemistry has typically reached steady state, is shown in Figure~\ref{fig:CH3OH_model-points} for \ce{CH3OH} gas and ice.

\begin{figure}
    \centering
    \includegraphics{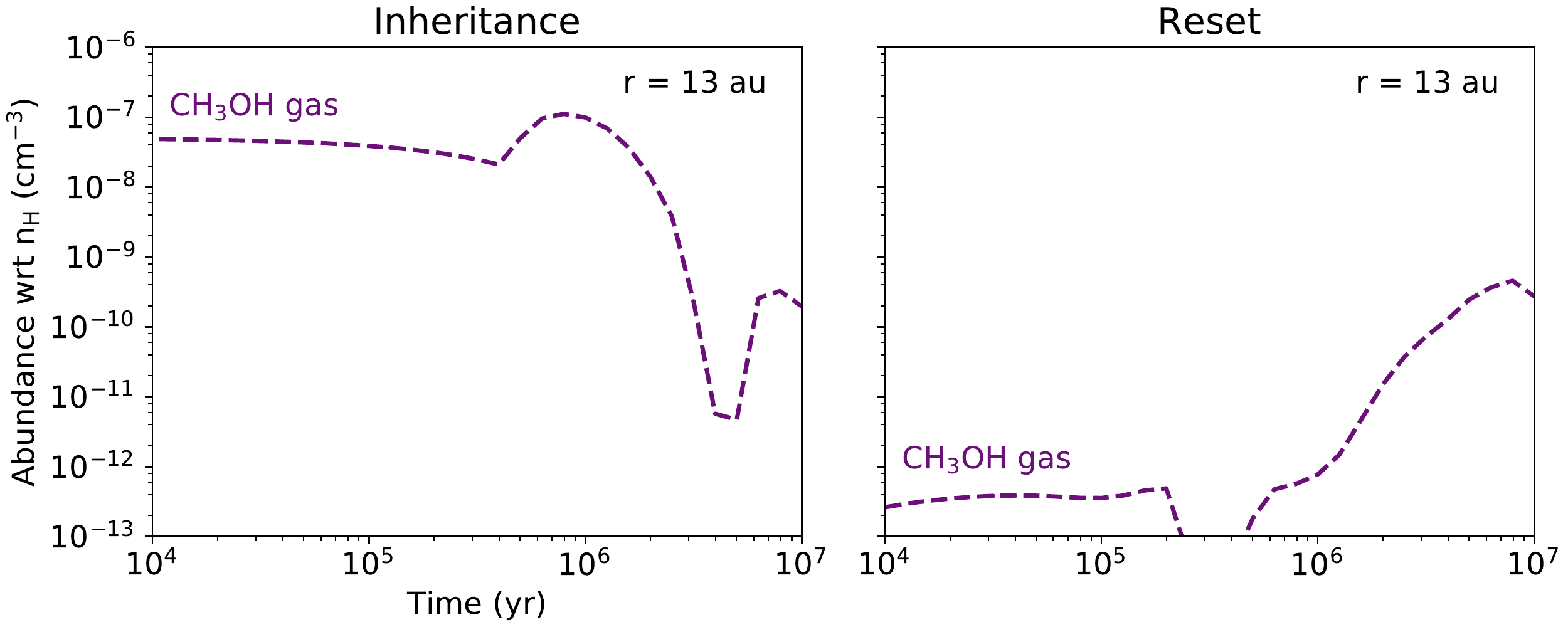}
    \hspace{3cm}
    \includegraphics{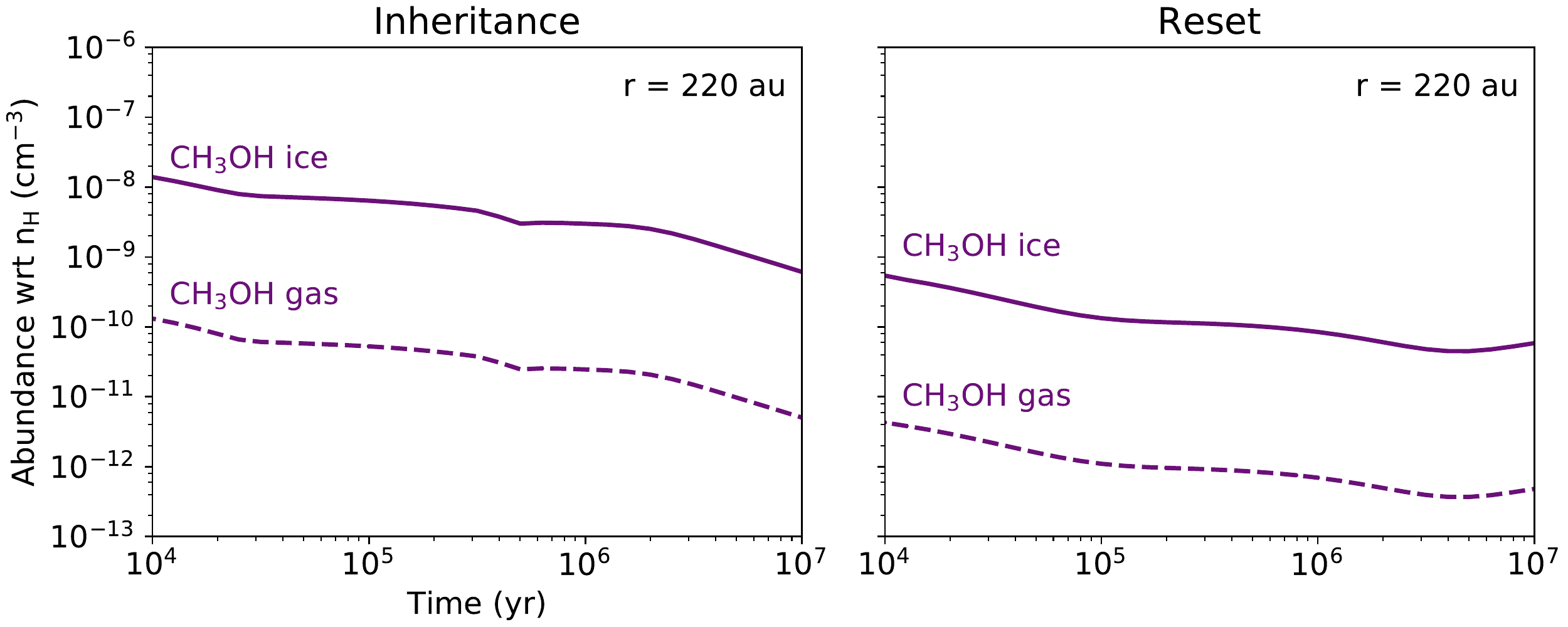}
    \caption{Fractional abundances of \ce{CH_3OH} at two points in the disk as a function of time. Right are with atomic ("chemical reset") initial conditions and left are with dark cloud ("chemical inheritance") initial conditions.}
    \label{fig:CH3OH_model-points2}
\end{figure}

The gas phase CO has the expected constant fractional abundance of $\approx 10^{-4}$ throughout the disk and there is very little CO ice which is consistent with 
\citep{2016A&A...588A.108K}. 
The \ce{H2CO} gas is in a layer in the disk atmosphere and there is only abundant \ce{CH3OH} gas ($\sim 10^{-7} - 10^{-6}$) in a vertical strip the inner disk at the edge of dust cavity.
The column densities of \ce{CH3OH} at 1~Myr, 2~Myrs and 5~Myrs are shown in Figure~\ref{fig:CH3OH_model-rad} and highlight that the \ce{CH3OH} abundance has not reached steady state by 5~Myrs. The chemistry was further investigated by looking at the change of \ce{CH3OH} in time at specific points in the disk: one at 13~au where the gas phase column density is highest and one at 220~au where 99\% of \ce{CH3OH} is on the ice. 
We investigated two scenarios, one where the initial abundances are molecular (i.e., inherited from an earlier cold phase as in the 2D models) and another where the initial abundances are all atomic. 
This latter case mimics complete chemical reset of the material in the disk.  
The results are shown in Figure~\ref{fig:CH3OH_model-points2}.
Under atomic initial conditions \ce{CH3OH} cannot form in the inner disk within 1~Myr and reaches fractional abundances no greater than $10^{-12}$.
In comparison, in the inherited model the gas phase \ce{CH3OH} can be sustained here for 1~Myr. 
The key reactions involved in the inherited model after instantaneous sublimation of the \ce{CH_3OH} ice are:
$$ \ce{CH3OH2+ + NH3 \rightarrow CH3OH + NH4+}.$$
and 
$$\ce{CH3OH + H3O+ \rightarrow CH3OH2+ + H2O.} $$
The \ce{CH3OH2+} ion primarily forms from proton transfer reactions with \ce{CH3OH}.
These gas phase reactions were originally found to be important in hot core chemistry after the sublimation of \ce{CH3OH} ice \citep{2016ApJ...821...46T}. 
This is the first example of such a phenomenon in a Class II protoplanetary disk.

In the "reset" model the increase in \ce{CH_3OH} at later times is also due to the above reaction but after 10~Myrs the abundance does not increase any further and is still $\approx 10^3 \times$ less than that required to match the observations. 
The increase in \ce{CH_3OH} at starting at 0.5 and up to 1.0 Myrs is due to this reaction:
    $$\ce{CH3O + HCO \rightarrow CH3OH + CO}~~~~~\Delta E = 0 \mathrm{K}$$

In the outer disk, at 220~au, there is no significant \ce{CH3OH} is formed in either model and this is primarily because the dust temperature is $> 20$~K and therefore there is no CO ice present for the hydrogenation reactions to proceed \citep{2002ApJ...571L.173W}. 

At early times
there is some formation of \ce{CH3OH} in the grains via 
$$\ce{OH_{\mathrm{ice}} + CH3_{\mathrm{ice}} \rightarrow CH3OH_{\mathrm{ice}}} ~\mathrm{or}~ \ce{CH3OH_{\mathrm{gas}}}$$
and 
$$\ce{OH_{\mathrm{ice}} + CH2_{\mathrm{ice}} \rightarrow CH2OH_{\mathrm{ice}} + GH  \rightarrow CH3OH_{\mathrm{ice}}}  ~\mathrm{or}~ \ce{CH3OH_{\mathrm{gas}}}.$$ 

but the destruction on the ice via hydrogenation on the grain surfaces and proton transfer in the gas phase via
$$\ce{HCO+ + CH3OH \rightarrow CH3OH2+ + CO} $$
are always more significant and the \ce{CH3OH} never increases above the initial inherited abundance.

\newpage

\renewcommand\refname{References}
\bibliography{xampl.bib}

\end{document}